# Phase Transitions In An Implicit Solvent Minimal Model Of Lipids: Role Of Head-Tail Size Ratio

**Biplab Bawali[1], Jayashree Saha[1], and Alokmay Datta[2]**
[1]Department of Physics, University of Calcutta, 92 Acharya Prafulla Chandra Ray Road, Kolkata 700 009
[2]Materials Characterization and Instrumentation Division, CSIR-Central Glass and Ceramic Research Institute, 196 Raja Subodh Chandra Mullick Road, Kolkata 700 032

*Corresponding author: *fellow1@cgcri.res.in; alokmay.datta@saha.ac.in; alokmaydatta@gmail.com*

**Abstract**

We present Monte Carlo simulations under constant NVT conditions on a minimal 'three beads' coarse grained implicit solvent model of lipid molecules, with the hydrophilic head represented by one bead and the hydrophobic tail represented by two beads. We consider two lipids, one with the head and tail bead sizes equal and the other with the tail beads smaller than the head. When cooled to the ambient temperature from an initial isotropic phase at high temperature, the first lipid transforms spontaneously to a lamellar phase while the second lipid transforms to a micellar phase, showing the crucial role of the head-tail size ratio on lipid phases.

## Introduction

Phase transitions in lipid assemblies have been at the centre of attention because of their association with biology and medicine [1-8]. Depending upon amphiphilic concentration, different phases like micellar, hexagonal, lamellar can be achieved in these systems. Other than lipid concentration, parameters like temperature, pH, the type of amphiphilic, water content and additives can change lipid phases.

From molecular structural perspectives, the lipid phase in a solvent depends on ratio of the size of head group to the aliphatic chain present in the amphiphilic lipid because the phase structures are dependent on the degree of curvature generated by the packing arrangement of the amphiphilic molecule. Low curvatures create lamellar phases whereas larger increment in curvature results in the formation of micellar phases.

The implicit solvent model accounts for the aqueous solvent by including its effects on the model lipid molecules. Here we present a minimal implicit solvent coarse grained model of lipids which is able to produce different phases by changing head to tail size ratio of the molecules.

## Model

In this model each lipid molecule consists of three spherical atoms (Figure 1). The blue sphere represents the hydrophilic head group while the two red spheres represent the hydrophobic tail group. This model is a modified version of the model proposed by Cooke and Deserno [9]. By eliminating the finite extensible nonlinear bonds between the beads we have fixed the bond length which makes the model simpler.

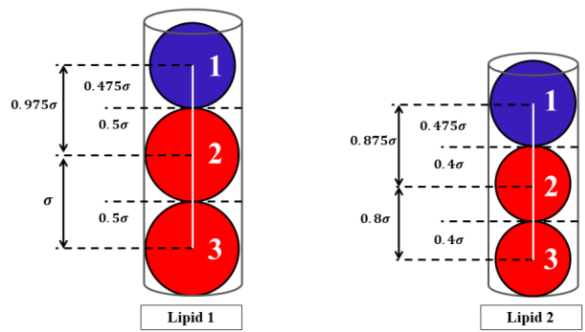

**Figure 1.** The three atom coarse grained lipid model. Lipid 1 has longer tail on the other hand Lipid 2 has smaller tail. Variation in the length of the tail chain is modeled by varying the diameter of the tail beads.

In this model, the effects of the solvent on each molecule have been considered implicitly in the interaction of the molecules. To incorporate the hydrophilic and hydrophobic effect each bead interacts with others via Weeks-Chandler-Andersen (WCA) repulsive potential

$$\phi(r;\sigma_{ij}) = \begin{cases} 4\varepsilon\left(\left(\frac{\sigma_{ij}}{r}\right)^{12} - \left(\frac{\sigma_{ij}}{r}\right)^{6}\right) + \varepsilon, & r \leq r_c \\ 0, & r > r_c \end{cases} \quad \text{a)}$$

To stabilize the lipid structure, an extra attractive potential between all tail beads are considered.

$$V_{tail-tail}(r) = \begin{cases} -\varepsilon & , r < r_c + w_f \\ 4\varepsilon\left[\left(\frac{\sigma_{ij}}{r-w_f}\right)^{12} - \left(\frac{\sigma_{ij}}{r-w_f}\right)^6\right] & , r_c \leq r \leq w_f + w_{cut} \\ 0 & , r > w_f + w_{cut} \end{cases} \quad b)$$

with $r_c = (2)^{1/6}\sigma$, where $\sigma_{ij}$ is the effective diameter of each atom. We have considered two lipids (Lipid-1 and Lipid-2). For Lipid-1, $\sigma_{ij}$ is chosen as $\sigma_{head,head} = 0.95\sigma, \sigma_{head,tail} = 0.975\sigma$, $\sigma_{tail,tail} = \sigma$ i.e. head and tails have same size. For Lipid-2, $\sigma_{head,head} = 0.95\sigma$, $\sigma_{head,tail} = 0.875\sigma$, $\sigma_{tail,tail} = 0.80\sigma$ i.e. tails are smaller than the head. Here $\sigma$ is the unit of length and $\varepsilon$ is the unit of energy. The value of $w_{cut} = 2.5\sigma$ and stabilization energy $w_f$ is chosen as $w_f = 0.4\sigma$. It should be mentioned that Lipid-2 may represent a different conformation of Lipid-1 with bond disorder in the tails.

## Result

We have a system of 800 lipid molecules within a cubic box of side $21.0\ \sigma$. We carried out Monte-Carlo (MC) simulations on the system for the two different types of lipid, under constant NVT conditions. First we have achieved the isotropic phase at a sufficiently high temperature for both lipids (Figure 2(a)). We then started lowering the temperature until it reached the reduced temperature 1.0, corresponding to ambient temperature. We then kept the temperature fixed and simulate both systems further with larger Monte-Carlo steps under identical conditions and without changing any other parameter.

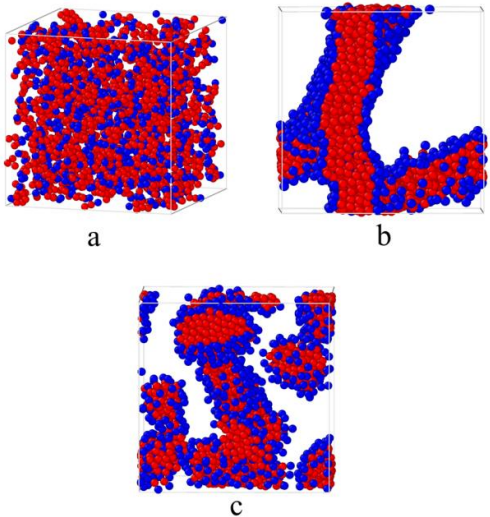

**Fig. 2.** Different phases of the system. (a) the isotropic phase (b) lamellar phase of Lipid 1 (c) micellar phase of Lipid 2.

The systems reduced to the lamellar phase for Lipid-1 (Figure 2(b)) and micellar phase for Lipid-2 (Figure 2(c)) after 8 lakh MC steps. No further change was observed.

## Conclusion

Lipid mesophases depend crucially on the head-tail size ratio of the lipid molecule. Here, for Lipid-1 the effective sizes of the head and tail beads are equal, which create zero mean curvature and gives rise to the lamellar phase. For Lipid-2 the head group is larger than the tail beads. This causes a positive large curvature and a micellar phase is observed in case of Lipid-2. Our minimal coarse-grained implicit solvent model of lipid molecules shows spontaneous phase transitions based entirely on the head-tail size ratio, underscoring this dependence.

## Acknowledgement

B.B gratefully acknowledges the support of Council of Scientific & Industrial Research (CSIR), India, for providing Senior Research Fellowship. A.D. acknowledges the Department of Atomic Energy for a Raja Ramanna Fellowship. OVITO software has been used to visualize the system.